\documentclass[aip,jap,showpacs,groupedaddress,reprint]{revtex4-1} 
\usepackage{amsmath} 
\usepackage{amsthm} 
\usepackage{amssymb}	
\usepackage{graphicx} 
\usepackage[usenames,dvipsnames]{pstricks}
\usepackage{pst-grad} 
\usepackage{pst-plot} 
\usepackage{color}
\usepackage{wasysym}
\usepackage{url}
\usepackage{bbold}
\usepackage{bm}
\newcommand{\markup}[1]{{\color{black}{#1}}}
\newcommand{\highlight}[1]{{\color{black}{#1}}}
\newcommand{\high}[1]{{\color{black}{#1}}}
\newcommand{\highl}[1]{{\color{black}{#1}}}
\newcommand{\highli}[1]{{\color{black}{#1}}}
\usepackage{subfigure}
\usepackage{gnuplottex}

\let\baraccent=\= 
\renewcommand{\=}[1]{\stackrel{#1}{=}} 

\theoremstyle{definition}

\theoremstyle{remark}


\begin{document}

\title{Photoluminescence transient study of surface defects in ZnO nanorods grown by chemical bath deposition}

\author{E. G. Barbagiovanni}
\email[]{eric.barbagiovanni@ct.infn.it}
\affiliation{MATIS IMM-CNR and Dipartimento di Fisica e Astronomia, Universit\`{a} di Catania, via S. Sofia 64, 95123 Catania, Italy}

\author{V. Strano}
\affiliation{MATIS IMM-CNR and Dipartimento di Fisica e Astronomia, Universit\`{a} di Catania, via S. Sofia 64, 95123 Catania, Italy}

\author{G. Franz\`{o}}
\affiliation{MATIS IMM-CNR and Dipartimento di Fisica e Astronomia, Universit\`{a} di Catania, via S. Sofia 64, 95123 Catania, Italy}

\author{I. Crupi}
\affiliation{MATIS IMM-CNR and Dipartimento di Fisica e Astronomia, Universit\`{a} di Catania, via S. Sofia 64, 95123 Catania, Italy}

\author{S. Mirabella}
\affiliation{MATIS IMM-CNR and Dipartimento di Fisica e Astronomia, Universit\`{a} di Catania, via S. Sofia 64, 95123 Catania, Italy}

\date{\today}

\begin{abstract}

Two deep level defects (2.25 and 2.03 eV) associated with oxygen vacancies (V$_o$) were identified in ZnO nanorods (NRs) grown by low cost chemical bath deposition. A transient behaviour in the photoluminescence (PL) intensity of the two V$_o$ states was found to be sensitive to the ambient environment and to NR post-growth treatment. The largest transient was found in samples dried on a hot plate with a PL intensity decay time, in air only, of 23 and 80 s for the 2.25 and 2.03 eV peaks, respectively. Resistance measurements under UV exposure exhibited a transient behaviour in full agreement with the PL transient indicating a clear role of atmospheric O$_2$ on the surface defect states. A model for surface defect transient behaviour due to band bending with respect to the Fermi level is proposed. The results have implications for a variety of sensing and photovoltaic applications of ZnO NRs. 

\end{abstract}


\maketitle


Zinc oxide is a wide \markup{band} gap ($\sim$ 3.2$\rightarrow$3.4 eV) n-type semiconductor with a large exciton binding energy (60 meV) \markup{and is a promising material for a range of} applications \cite{Janotti:2009}. However, \high{studies on low cost ZnO nanostructures (NSs) and thin films are unclear as to the} source of n-type conductivity \high{and persistent photoconductivity (PPC)}, the UV sensing mechanism, and the defect landscape \cite{Spencer:2013, Janotti:2007}. \markup{In particular, \highl{the correlation between the defect landscape and sensing (whether gas, pH, or UV) responsivity in low cost ZnO NRs is debated.}} In one model, the neutral oxygen vacancy (V$_o$) is an n-type donor state \cite{Janotti:2007}, atmospheric O$_2$ absorbs at this site \cite{Spencer:2013}, and a depletion \high{region forms beneath the surface} \cite{Liu:2010_1}. UV excitation \high{creates electron-hole pairs,} holes migrate to the depletion \highlight{region} and O$_2$ desorption occurs, thus reducing the depletion region and increasing the conductivity \cite{Liu:2010_1, Kushwaha:2012}. \highlight{Therefore,} the kinetics of O$_2$ desorption determine the response time of \highlight{a} UV sensor. In a second model, V$_o$ is reported to be a deep level state (DLS) and so cannot \highlight{be a donor state} \cite{Janotti:2009}. Instead, H occupying \highlight{V$_o$} sites act as donor states \cite{Spencer:2013, Janotti:2009}, therefore, the O desorption model does not drive UV sensing. An alternative \highlight{model} suggests that after UV excitation, the DLS \highlight{forms} a metastable state resonant with the conduction band (CB), \highlight{while} the doubly ionized V$_o$ (V$_o^{2+}$) state \highlight{lies} above the CB \cite{Lany:2005}. This mechanism is reported to explain PPC in UV sensors \cite{Hullavarad:2009, Spencer:2013}. Nonetheless, serious critiques over the validity of this model \high{have} been presented in the literature \cite{Janotti:2007}. 

In either model, the energetic position \highlight{of the ionized V$_o$ states} are central. Experimentally, defect states depend \highlight{on} the carrier concentration and hence the Fermi level (E$_F$) \cite{Wang:2012_2}. A general consensus finds that V$_o$ lies within the flat band region, while the depletion region is dominated by the singly ionized V$_o$ (V$_o^{+}$) and/ or V$_o^{2+}$ for ZnO \high{NSs} \cite{Wang:2012_2, Cheng:2013, Bouzid:2009, Chaudhuri:2010, Kushwaha:2012}. Theoretically, \highlight{the defect energy} depends on the chemical potential and lattice relaxations \cite{Janotti:2007, Lany:2005}, which vary between bulk materials, thin films, and NSs \cite{Janotti:2009}. To understand the role of the defect states with respect to O$_2$ desorption, we measured, \high{in air} and vacuum conditions, the defect photoluminescence (PL) peak \highlight{intensity} transient over large time scales \highlight{in ZnO nanorods (NRs)}. We \highlight{found a striking correlation between} the change in PL \highlight{intensity} \high{and} \highlight{NR} resistance under UV excitation. \high{These results indicate a transient in the surface defect structure, whereby, we present a unified model for O$_2$ desorption.}


ZnO NRs were grown by chemical bath deposition (CBD). A Si substrate with native oxide was seeded with a ZnO nanoparticle \high{solution (0.1 wt$\%$ in ethanol)} via spin-coating \highl{(200 rpm, 10s and 700 rpm, 20 s)}. The seeded substrate was then placed in a solution of 25 mM zinc nitrate hexahydrate [Zn(NO$_3$)$_2 \cdot$6H$_2$O] and 25 mM hexamethylenetetramine [C$_6$H$_{12}$N$_4$] (HMTA) at a \markup{temperature of} 90 $^o$C \cite{Strano:2014}. The quality, diameter, and length of the NRs was measured using \high{a Gemini field emission scanning electron microscopy (SEM) Carl Zeiss SUPRATM 25}. CBD introduces water based absorbates \cite{Tam:2006, Wagata:2012, Bera:2009}, we controlled this parameter by comparing samples as-prepared not dried (ND), dried at 100 $^o$C for 20 min on a hot plate, \highlight{and left to dry over two weeks}. Additionally, we annealed a sample \high{at 600 $^o$C for 30 min in O$_2$} to understand the role of V$_o$. 
 
Resistance measurements were performed under exposure to 364 nm UV light. The samples were biased to force a current of 1 nA \high{between two probes 1 mm apart to extract the resistance}. This value was used to avoid \highl{compliance} in the samples and gave a good overall response \highlight{under} UV exposure. \markup{PL measurements were performed by pumping at 1.5 mW the 325 nm line of a He-Cd laser chopped through an acousto-optic modulator at a frequency of 55 Hz. The PL signal was analyzed by a single grating monochromator, detected with a Hamamatsu visible photomultiplier, and recorded with a lock-in amplifier using the acousto-optic modulator frequency as a reference.} \high{PL spectra were} taken in air \high{or} vacuum \high{within a} cryostat ($\sim$10$^{-6}$ mbar), to ascertain the role of \highlight{atmospheric} O$_2$. \high{The PL transient behaviour was measured by fixing the monochromator at each peak wavelength and monitoring the PL intensity as a function of time.}
	
\begin{figure}
\includegraphics[scale=.32]{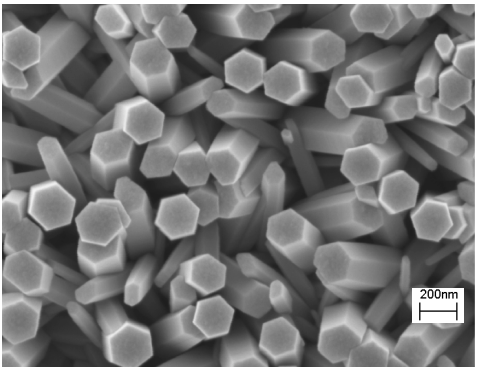} \caption{SEM image of the ZnO NRs showing their hexagonal crystal structure. \markup{Ref. \onlinecite{Strano:2014} measured their wurtzite \high{crystal} structure.} \label{SEM_ZnO}}
\end{figure}

An SEM image of the ZnO NRs is given in Fig. \ref{SEM_ZnO}. Lower resolution \highlight{SEM} images \high{(not shown)} demonstrate uniform coverage of the NRs over \highlight{a large} sample area. \high{Our CBD} ZnO NRs have a diameter and height $\sim$150 nm and 1 $\mu$m, respectively. The NRs \highl{are} oriented along the ZnO c-axis \cite{Strano:2014}, \highl{and} \markup{perpendicular to the substrate surface}. \high{The NR structure remains} in the annealed sample, though \highl{a few} nano-pits formed on the tips of the NRs. 

\begin{figure}
\includegraphics[scale=.635]{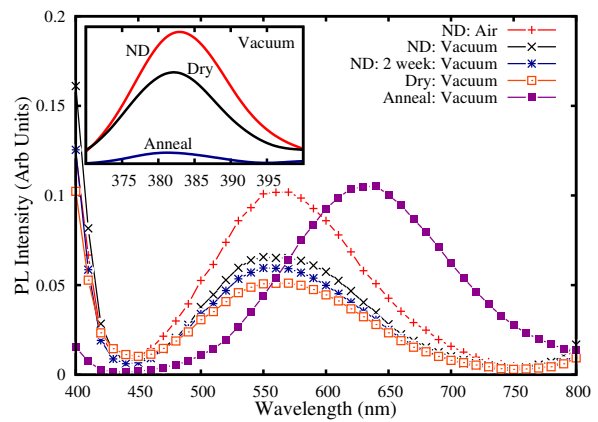} \caption{PL spectra showing the variation in the \high{DLS} peak intensity between air and vacuum for the dry, ND, and annealed samples. The inset \high{shows} the UV spectra for the dry, ND, and annealed samples in vacuum.\label{Defect_PL}}
\end{figure}

\high{The effect of drying and O$_2$ annealing on the PL spectra is demonstrated in Fig. \ref{Defect_PL}. The spectra consists of two peaks, one in the UV (shown in the inset of Fig. \ref{Defect_PL}) and one at $\approx$ 550 nm related to defects.} \high{The defect peak, in all samples, can be fitted with two DLS peaks at 555 nm (2.25 eV) and 610 nm (2.03 eV). The annealed sample exhibits a large red shift of the defect peak to $\approx$ 650 nm, which can be fitted with two DLS peaks at 610 and 651 nm, and has the largest PL intensity in vacuum indicating we may have increased the Zn-related defect concentration \cite{Borseth:2006}.} \highlight{Therefore, due to the reduction of the 555 and 610 nm peaks after annealing, we associate these peaks with V$_o$ \cite{Kushwaha:2012, Borseth:2006}.} \highlight{Furthermore, Fig. \ref{Defect_PL} \highlight{demonstrates} a decrease \high{of} the \highlight{DLS} peak intensity from the ND to the 2 week \high{aged} and the dried sample. This feature is understood as a result of organic species \cite{Wagata:2012}, defect complexes \cite{Studenikin:2002} due to water \cite{Bera:2009}, or OH groups \cite{Tam:2006} loosely bound to the \highlight{NR} surface introduced during CBD. These defect species can act as shallow donors, are easily removed \high{after} UV exposure \cite{Wagata:2012}, and create a $\pm$10 nm shift in the DLS PL spectrum. Therefore, we conclude that drying the sample removes these defect species from the surface lowering, \high{but stabilizing,} the PL intensity.} The inset in Fig. \ref{Defect_PL} shows the UV peak centred \high{at} 382 nm \highlight{for all samples}. The intensity of the ND sample is almost 1.5 times the dried sample, and 7.8 times the annealed sample. This result is likely \high{due to the observation that} transitions do not \high{occur} from band to band, but with shallow defect states near the conduction band minimum (CBM) or \high{valence band maximum} (VBM), which \high{are} reduced during the O$_2$ anneal \cite{Srikant:1998}. Many authors state \highlight{that a reduction in the UV peak} indicates a reduction in the optical quality of the sample, and thus a reduction of the UV sensing capability \cite{Kushwaha:2012, Liu:2010_1}. We believe that this assessment is premature as it does not segregate the role \highlight{of different defects types in UV sensing.}

\begin{figure}
\includegraphics[scale=.635]{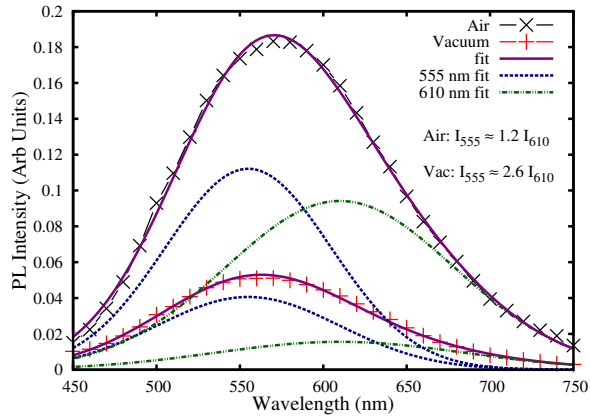} \caption{DLS PL spectrum for the dried sample in vacuum and air.\label{Defect_air_vac}}
\end{figure}

\high{Fig. \ref{Defect_air_vac} shows the PL spectra for the dried sample taken both in air and vacuum.} \high{The DLS PL intensity is higher when measured in air}, which is consistent with the \markup{ND} sample (see Fig. \ref{Defect_PL}). \highlight{This result indicates that atmospheric O$_2$ promotes defect \highl{radiative} transitions, which we discuss further below. \high{The 555 and 610 nm fits are shown,} with a relative intensity of I$_{555}\approx$ 1.2 I$_{610}$ and I$_{555}\approx$ 2.6 I$_{610}$ in air and vacuum, respectively. We ascribe the 555 nm and 610 nm peaks to surface related V$_o^{+}$, and V$_o$\high{\cite{Tam:2006}}, respectively.} \highli{Though our assignment of the V$_o$ states is indirect, it is in agreement with the literature \cite{Tam:2006, Borseth:2006, Vlasenko:2006}, and supported by our model below.}

\begin{figure}
\includegraphics[scale=.635]{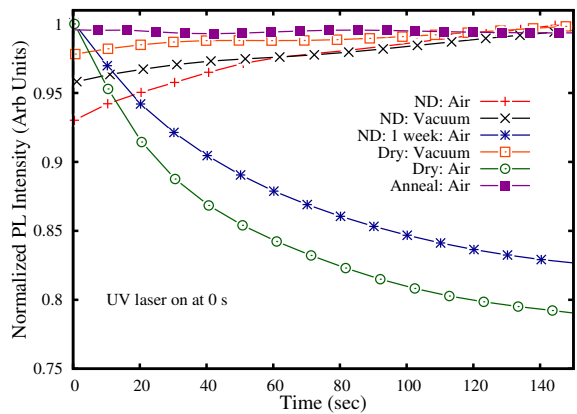} \caption{555 nm PL transient in air and in vacuum for the dried, ND, \markup{aged} and annealed sample.\label{PL_time}}
\end{figure}

\high{We measured the 555 nm PL intensity as a function of time to understand the effect of UV excitation on our ZnO NRs, shown in Fig. \ref{PL_time}.} First, the ND sample \highlight{exhibits} essentially no change in the PL intensity whether in air or vacuum. \highlight{We assume the PL transient is suppressed because of the loosely bound surface defect states.} \high{In contrast, after 1 week of drying, we measured a PL transient with an exponential decay time} of $\sim$ 70 s. Comparing with the dried sample, in vacuum there is no transient, while in air there is a marked \high{exponential decay with two components at 23 and 165 s, \highl{and a relative intensity of I$_{23s}\approx$ 1.2 I$_{165s}$}}. On the other hand, the annealed sample shows no defect transient, because of the reduction in the V$_o$ concentration. A similar result was found in UV \highlight{sensors}, whereby, O$_2$ annealing reduces the UV sensing performance \cite{Kushwaha:2013, Lv:2013}. \markup{Furthermore, the transient time of the 610 nm peak \high{(not shown)} is slower than the 555 nm peak \high{with a single component $\approx$} 80 s.}

\begin{figure}
\includegraphics[scale=.635]{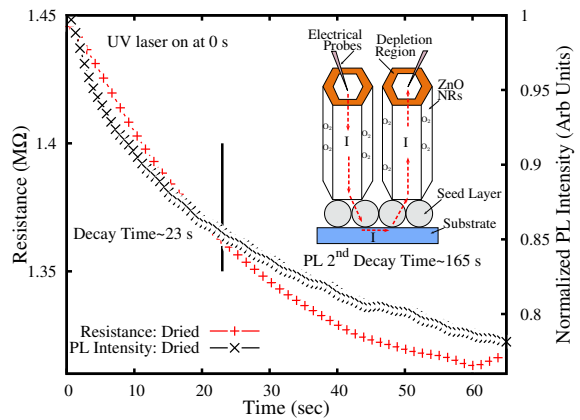} 
\caption{Comparison of the resistance (2$\mu$/cm$^2$, 364nm UV light) and 555 nm PL transient for the dried sample. Both transients have a decay time of 23 s, while the 2$^{nd}$ decay time is for the PL transient alone. \highli{The insest represents the resistance measurement geometry.} \label{Res_PL}}
\end{figure}

Finally, \high{in Fig. \ref{Res_PL}, a comparison between sample resistance \highlight{and PL intensity} as a function of time under UV exposure is shown for the dried sample.} \highli{The inset of Fig. \ref{Res_PL} depicts the geometry of the electrical probes during the resistance measurements. The electrical probes force a current along the NR length perpendicular to the depletion extension.} There are several factors that can affect the transient measurements, such as the intensity of the excitation source and sample quality. Nonetheless, \highlight{we consistently measure an \high{exponential} decay time in the resistance $\sim$ 20$\rightarrow$25 s, which is well correlated with the first decay time of the 555 nm \highl{PL} peak. \highl{Resistance} measurement using a Au interdigitated mask grown atop the Si substrate prior to ZnO NR growth \highl{gives} faster decay time of 5 s (not shown). While, in the annealed sample (not shown) we measured no variation in the resistance under UV exposure.} We can explain these results with a simple model. 

\begin{figure}
\includegraphics[scale=.85]{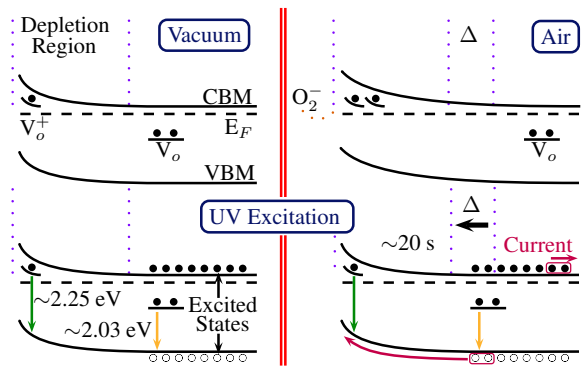}
\caption{Schematic illustration of the UV sensing mechanism (not drawn to scale). The top and bottom half represent the band configuration in air and vacuum before and after UV excitation, respectively. Band bending ionizes the V$_o$ state at the surface to V$_o^{+}$. In the air configuration, absorbed O$_2$ increases the band bending (represented by $\Delta$) and the number of V$_o^{+}$ sites. After UV excitation, two DLS PL bands are represented. In air, holes (open circles) migrate to the surface allowing O$_2$ to desorb decreasing the depletion region and the number of V$_o^{+}$ sites, simultaneously electron (filled circles) conduction increases. \label{band}}
\end{figure}

The results \high{reported} above hint at a \highl{clear} role for surface defects under UV excitation, depicted in Fig. \ref{band}. \markup{First,} the depletion region \highlight{results from} \high{accumulated electrons at the surface due to broken lattice order \cite{Cheng:2013}.} \markup{In the flat band region, V$_o$ is neutral (doubly occupied) \cite{Janotti:2007} below E$_F$ \cite{Studenikin:2002}. Upward band bending in the depletion region forces V$_o$ above E$_F$, which ionizes the state to become V$_o^+$ (singly occupied) \cite{Studenikin:2002, Wang:2012_2} (see Fig. \ref{band} `vacuum'). Atmospheric O$_2$ absorbs on the surface at the V$_o^+$ sites by \highl{polarizing} surface electrons and forming O$_2^-$, thus further increasing the depletion region \cite{Wang:2012_2}, represented by $\Delta$ in Fig. \ref{band} `air'. This picture differs from what was \highl{reported earlier}, because atmospheric O$_2$ does not initially attach to conduction electrons. It is also possible that V$_o^{2+}$ (unoccupied) states will form at the surface with sufficient band bending \cite{Bouzid:2009}. At the same time, possible trap-filling at the surface reduces the ionization state of V$_o^{2+}$ or V$_o^+$ \highl{(Ref. \onlinecite{Cheng:2013})} and stabilizes the band bending, hence, we assume that the surface is mainly comprised of V$_o^{+}$ sites \cite{Chaudhuri:2010, Kushwaha:2012}. \highlight{Therefore, in air, we expect more V$_o^{+}$ and V$_o$ PL centres with a greater probability of transition, because of the larger depletion region.} This picture explains why, in air, the defect PL intensity is larger and I$_{555}\approx$ I$_{610}$.}

The bottom part of Fig. \ref{band} depicts what happens after UV exposure. The UV excitation creates an occupation of excited electrons and holes, given by the closed and open circles in Fig. \ref{band}, respectively. In vacuum, there is a PL spectra from the V$_o^+$ and V$_o$ state at 2.25 and 2.03 eV, respectively, which is constant over time. While in air, the holes are free to migrate to surface and neutralize O$_2^-$, allowing O$_2$ to desorb from the sample surface, \markup{which reduces the depletion region and induces two phenomena. First, electrons in the CB are free to conduct and thus a decrease in resistance is measured. Second, as the depletion region bends down the ionization state of V$_o^+$ ($\rightarrow$ V$_o$) reduces, thus reducing the number of radiative centres, which explains why the PL intensity decreases over time and why the 610 nm peak decays slower than the 555 nm peak. \highlight{Therefore, O$_2$ desorption drives both the PL and resistance transient over the time scale of \high{$\sim$20 s in our samples}, while the longer PL transient may be associated \high{with the} variation in the V$_o$ ionization state and requires further investigation.}}

This model of surface defects can help to explain some of the conflicting results in the literature. For example, it has been argued that O$_2$ desorption is not a significant process in PPC studies \cite{Hullavarad:2009, Spencer:2013}, whereby, the metastable state model of Ref. \onlinecite{Lany:2005} is favoured. However, these studies consider thin films \cite{Spencer:2013, Li:2005}, which have a lower concentration of surface defects and thus a modified O$_2$ desorption rate compared with NRs \cite{Bayan:2012, Li:2005, Swanwick:2012, Bera:2009}. In our model, since conduction electrons accumulate at the surface V$_o^+$ state, response time is correlated with V$_o^+$ due to surface band bending. Therefore, more band bending implies more V$_o^+$ states and promotes the V$_o^+$ $\rightarrow$ V$_o$ transition, which can explain why better response was observed for gas sensors with reduced NR diameter \cite{Lupan:2010}. Furthermore, our model implies O$_2$ desorption is only one route to obtain charge separation through the V$_o^+$ state. Many authors have found enhanced UV sensing by coating ZnO with a metal \cite{Liu:2010_1} or a conducting polymer \cite{Hassan:2012}, in agreement with our results using a Au interdigitated mask.  


In conclusion, our low cost CBD ZnO NRs exhibit two main surface DLS peaks at 555, and 610 nm due to V$_o^{+}$, and V$_o$, respectively. The 555 nm PL intensity exhibits a transient in air $\sim$ 20 s, which is well correlated with the change in resistance under UV excitation. This correlation arises because O$_2$ desorption decreases the band bending and thus the concentration of V$_o^{+}$ states, simultaneously, charge separation reduces the sample resistance. The PL transient is suppressed in vacuum, because the depletion region is stable since O$_2$ desorption does not occur. We presented a unified model for these results with implications for photovoltaic, gas, pH, and UV sensing applications.

%

\acknowledgements
We would like to thank Kingsley Iwu for valuable discussions and insights regarding the chemical synthesis of ZnO NRs. The authors acknowledge MIUR projects, ENERGETIC (PON02$\textunderscore$00355$\textunderscore$3391233), and PLAST$\textunderscore$ICs (PON02$\textunderscore$00355$\textunderscore$3416798).

\end{document}